\documentclass[aps,prl,superscriptaddress,showpacs,preprintnumbers,nofootinbib,reprint,floatfix]{revtex4-1}
\usepackage{mciteplus}
\usepackage{latexsym}
\usepackage{amsthm}
\usepackage{amsmath}
\usepackage{amssymb}
\usepackage{hepunits}
\usepackage[colorlinks]{hyperref}
\hypersetup{
citecolor=red
}
\usepackage{bbm}
\usepackage{bm}
\usepackage{xfrac}
\usepackage{color}
\usepackage{colordvi}
\usepackage{comment}
\usepackage{dcolumn}
\usepackage{times,latexsym,graphicx,wrapfig,url}
\usepackage{epsfig,lineno,bm}
\usepackage{subfigure}
\usepackage{slashed}
\usepackage[normalem]{ulem}
\usepackage{xcolor} 
\usepackage{wasysym}
\allowdisplaybreaks

\def\ie{{\it i.e.}}
\def\hc{{\rm h.c.}}
\def\tr{{\rm Tr}}

\usepackage{calrsfs}
\DeclareMathAlphabet{\pazocal}{OMS}{zplm}{m}{n}

\newcommand{\bil}[1]{{#1^\dagger #1}}
\newcommand{\bils}[1]{{#1^* #1}}
\newcommand{\vev}[1]{{\langle #1\rangle}}

\newcommand{\be}{\begin{eqnarray}}
\newcommand{\ee}{\end{eqnarray}}
\newcommand{\bea}{\begin{eqnarray}}
\newcommand{\eea}{\end{eqnarray}}

\newcommand{\bef}{\begin{figure}[htbp]\begin{center}}
\newcommand{\eef}{\end{center}\end{figure}}
\newcommand\FNAL{Fermi National Accelerator Laboratory, Batavia, IL, 60510, USA}
\newcommand\USP{Departamento de F\'{i}sica Matem\'{a}tica, Instituto de F\'{i}sica, Universidade de S\~{a}o Paulo, Rua do Mat\~{a}o 1371, CEP. 05508-090 S\~{a}o Paulo, Brazil}
\newcommand\UAM{Departamento  de  F\'{\i}sica Te\'{o}rica,  Universidad  Aut\'{o}noma  de  Madrid,
\it Cantoblanco  E-28049  Madrid,  Spain}
\newcommand\IFT{Instituto  de  F\'{\i}sica  Te\'{o}rica  UAM/CSIC,
 Calle Nicol\'{a}s Cabrera  13-15,  Cantoblanco  E-28049  Madrid,  Spain}
\newcommand\PITT{Department of Physics and Astronomy, University of Pittsburgh, 3941 O'Hara St., Pittsburgh, PA 15260, USA}
\newcommand\IFTUNESP{ICTP South American Institute for Fundamental Research \& Instituto de F\'{i}sica Te\'orica, UNESP, Rua Dr.\ Bento T.\ Ferraz 271, CEP.\ 01140-070, S\~{a}o Paulo, Brazil}
\def\lsim{\mathrel{\rlap{\lower4pt\hbox{\hskip1pt$\sim$}}
    \raise1pt\hbox{$<$}}}
\def\gsim{\mathrel{\rlap{\lower4pt\hbox{\hskip1pt$\sim$}}
    \raise1pt\hbox{$>$}}} 
    
\graphicspath{{Figs/}}

\begin{document}

\preprint{FTUAM-18-10}
\preprint{PITT-PACC-1809}
\preprint{FERMILAB-PUB-18-127-T}

\title{Natural and Dynamical Neutrino Mass Mechanism at the LHC}

\author{Julia Gehrlein}          \thanks{julia.gehrlein@uam.es}  \affiliation{\UAM} \affiliation{\IFT} 
\author{Dorival Gon\c{c}alves}          \thanks{dorival.goncalves@pitt.edu}  \affiliation{\PITT} 
\author{Pedro~A.~N.~Machado}          \thanks{pmachado@fnal.gov; ORCID: 0000-0002-9118-7354}  \affiliation{\FNAL}
\author{Yuber F. Perez-Gonzalez}          \thanks{yfperez@if.usp.br}  \affiliation{\USP} \affiliation{\IFTUNESP} 

\date{\today}
\begin{abstract}
We generalize the scalar triplet neutrino mass model, the type II seesaw. Requiring fine-tuning and arbitrarily small parameters to be 
absent leads to dynamical lepton number breaking at the electroweak scale and a rich LHC phenomenology. A smoking gun signature 
at the LHC that allows to distinguish our model from the usual type II seesaw scenario is identified. Besides, we discuss other interesting 
phenomenological aspects of the model such as  the presence of a massless Goldstone boson and deviations of standard model Higgs couplings.
\end{abstract}

\maketitle

\noindent
{\bf I. Introduction}\\
The presence of non-zero neutrino masses, as inferred by neutrino oscillation experiments, is the only laboratory-based e\-vi\-den\-ce 
of physics beyond the standard model~\cite{Kajita:2016cak, McDonald:2016ixn}. Strictly speaking, neutrinos have no mass in the 
standard model (SM). There is no unique prescription of how neutrino could become massive. Perhaps the simplest way of generating 
neutrino masses is via the seesaw framework. In its na\"{i}ve rea\-li\-za\-tions, seesaw types I, II and III \cite{Mohapatra:1986bd, GellMann:1980vs, Mohapatra:1979ia, Mohapatra:1980yp, Yanagida:1979as, Schechter:1980gr, Lazarides:1980nt}, a large suppression of the electroweak 
breaking scale provides an explanation for the smallness of neutrino masses. Without a full underlying framework, like Grand Unified 
Theories or Supersymmetry, these mechanisms typically introduce a hierarchy problem due to the large mass gap~\cite{Vissani:1997ys} 
or rely on very small (but technically natural~\cite{tHooft:1979rat}) parameters.

In general, the seesaw mechanism generates a small pa\-ra\-me\-ter from the ratio of two disparate physics scales, e.g., electroweak versus 
Grand Unification scales. Therefore, when we set the new heavy states to the weak scale (such as done in studies of type II seesaw at 
colliders~\cite{Perez:2008ha, Perez:2008zc}), the ``seesaw'' mechanism is exchanged by a small parameter. This can be appreciated in a 
model independent way by writing down sche\-ma\-ti\-cal\-ly the Weinberg effective operator that generates neutrino 
masses~\cite{Weinberg:1979sa}, 
namely
\begin{equation}
\mathcal{L}_5=\frac{c}{\Lambda}LLHH
\end{equation}
($H$ and $L$ are the Higgs and lepton doublets)
and observing that if $\Lambda\sim\langle H\rangle$ then the Wilson coefficient $c$ needs to be tiny in order to obtain sub-electronvolt 
neutrino masses. We will show in this Letter that a simple generalization of the type II seesaw can dynamically generate this small parameter 
by replacing the seesaw by a chain of seesaws.

\begin{figure}[t]
\centering
\includegraphics[width=0.55\columnwidth]{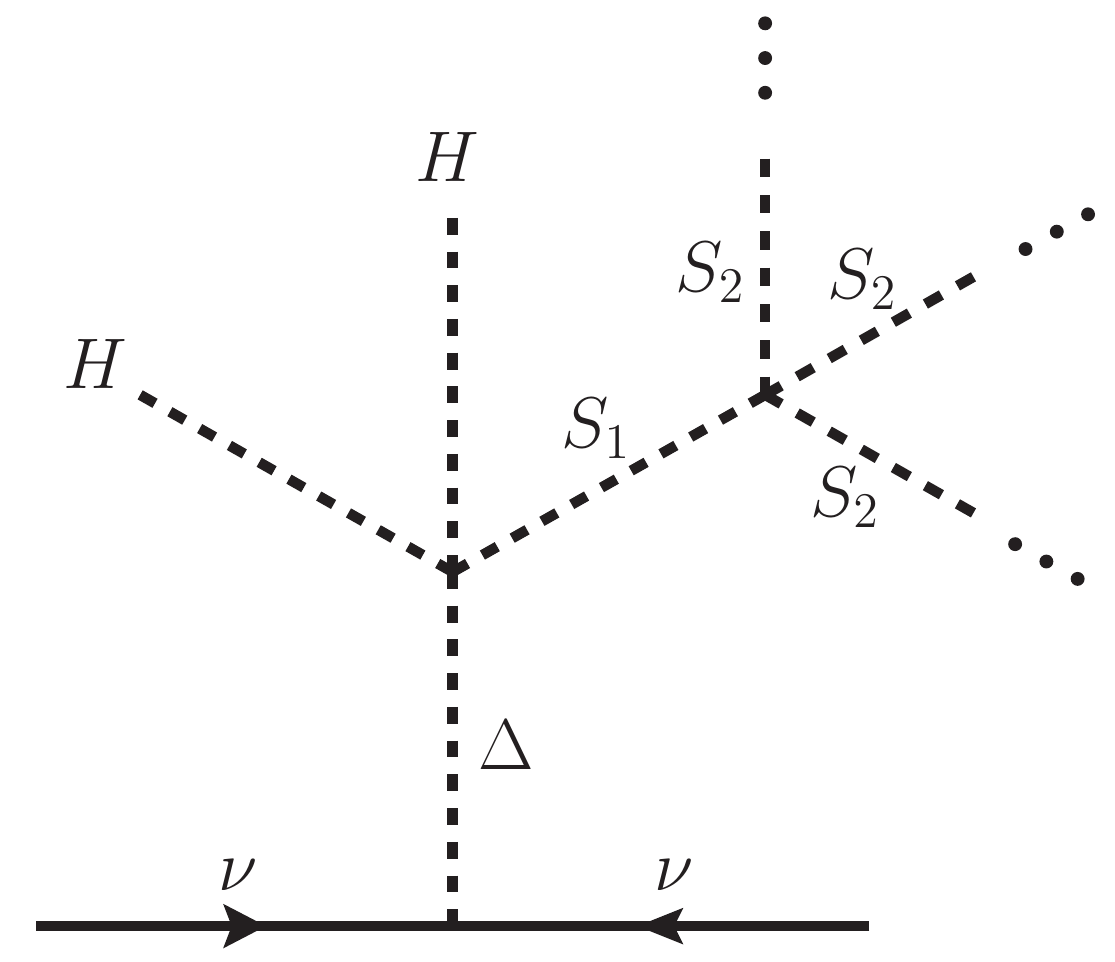}
\caption{Illustration of the generalized type II seesaw mechanism for neutrino mass generation.}
 \label{fig:model}
\end{figure}

More concretely, in type II seesaw a scalar triplet 
\begin{equation}
\Delta=\left( \begin{array}{cc} \delta^+/\sqrt{2}& \delta^{++}\\ (v_\Delta+\delta+ia_\delta)/\sqrt{2} &-\delta^+/\sqrt{2}\end{array}\right)
\end{equation}
obtains its vacuum expectation value (vev) after electroweak symmetry breaking
\begin{equation}
  v_\Delta\simeq \frac{\mu}{\sqrt{2}} \frac{v^2}{M_\Delta^2},
\end{equation}
where $\mu$ is a dimensionful lepton number breaking parameter of the scalar potential, $v=246~\GeV$ is the Higgs doublet vev, 
and $M_\Delta$ is approximately the physical mass of $\Delta$. Neutrino masses are given by $m_\nu= \sqrt{2}Yv_\Delta$, with $Y$ 
being a matrix of Yukawa couplings.

We can immediately see that the smallness of neutrino masses can only be obtained by having small Yukawas, large $M_\Delta$, and/or 
small \emph{ad hoc} lepton number breaking parameter $\mu$. For instance, if $M_\Delta$ is accessible at the LHC, say at the TeV scale, 
and the Yukawas are taken to be of order 1, we obtain
\begin{equation}
  \mu\simeq 1.6~ {\rm eV}~\left(\frac{m_\nu}{0.1~{\rm eV}}\right).
\end{equation}
Since  $\mu=0$ restores a symmetry of the Lagrangian, it is not generated by other couplings due to quantum corrections, thus being technically 
natural in the t'Hooft sense~\cite{tHooft:1979rat}. Nevertheless, it is unappealing to have this enormous hierarchy of scales 
$\mu/v\lesssim \mathcal{O}(10^{-11})$ put in arbitrarily. As suggested by the considerations made before regarding the Weinberg operator, 
this is not exclusive to type II seesaw.

In this Letter we present a generalization of the type II seesaw scenario which dynamically generates a very low lepton number breaking scale 
from a small hierarchy. The model is naturally found at the weak scale, introducing no new fine-tuning neither arbitrarily small couplings. Our 
mechanism engenders a rich and vast phenomenology, including deviations of SM Higgs couplings, the presence of a massless Majoron, lepton 
flavor violation and a smoking gun signature at the LHC which allows to distinguish this model from the usual type II seesaw.

\vskip 0.1cm
\noindent
{\bf II. The mechanism} \\
The idea simply amounts to  replicate the induced vev suppression mechanism with additional scalar singlets, as shown in Fig.~\ref{fig:model}. 
In our concrete setup, all mass parameters are  near the electroweak scale and all dimensionless couplings are of similar order, thus yielding a 
natural model of neutrino masses accessible at the LHC. We will focus on a scenario with two extra scalar singlets, as this is the most minimal 
realization that successfully implements the mechanism and also exhibits all important phenomenological features of our framework.
 
First   we require dynamical lepton number breaking. To that end, we promote $U(1)_\ell$ lepton number to a  global symmetry in which leptons 
have charge $\ell_{\rm leptons}=+1/2$ (the nor\-ma\-li\-za\-tion has been chosen for convenience) and quarks have no charge. The neutrino 
Yukawa coupling
\begin{equation}\label{eq:neutrino-yukawa}
  \mathcal{L}_{\rm Yuk}^\nu = -Y(L^c)^T i\sigma_2\Delta\, L + \hc
\end{equation}
($\sigma_2$ is the second Pauli matrix, $Y$ is a matrix of Yukawa couplings in flavor space, and $^c$ denotes charge conjugation) requires 
$\ell_\Delta=-1$, forbidding the triple coupling $\mu H^T i\sigma_2\Delta^\dagger H$. We introduce the first complex SM singlet scalar 
$S_1$ with lepton number $\ell_1=+1$ so its vev may play the role of the lepton number violating parameter $\mu$. Then, we generalize the 
type II seesaw model by invoking another extra scalar singlet with charge $\ell_2=1/3$, allowing for a term $S_{1}^* S_{2}^3$ in the scalar 
potential. All scalars but the Higgs and $S_2$ have positive bare mass terms.The crucial point is that when $S_{2}$ develops a vev spontaneously, 
it induces a suppressed vev for $S_{1}$, which then induces an even smaller vev for $\Delta$. The model can easily be ge\-ne\-ra\-li\-zed 
for any number $N$ of scalar singlets, see Appendix~A. We identify the usual type II seesaw with a $N\!=\!1$-step version of the generalized 
model in which $S_1$ is integrated out. Our model bears similarities with multiple seesaw and clockwork models (see, for instance, 
Refs.~\cite{Dudas:2002ry, Xing:2009hx, Bonilla:2015jdf, Ishida:2017ehu, Gu:2017gra, Choi:2014rja, Choi:2015fiu, Kaplan:2015fuy, Giudice:2016yja}).

As we will see later, a simple 2-step realization can lead to small neutrino masses given that some quartic couplings and neutrino Yukawas 
are of order $10^{-2}\sim 10^{-3}$ (larger couplings can be obtained in realizations with extra steps).
Without further ado, we write down the scalar potential 
\begin{widetext}
\begin{align}
  V= &-\frac{m_H^2}{2}\bil{H}  +m_\Delta^2\vev{\bil{\Delta}} + m^2_1\bils{S_1}-\frac{m^2_2}{2}\bils{S_2} + \frac{\lambda_H}{4}(\bil{H})^2+\frac{\lambda_2}{4}(\bils{S_2})^2 \nonumber\\
 & + \lambda_{1H}(\bils{S_1})(\bil{H}) + \lambda_{2H}(\bils{S_2})(\bil{H}) + \left[ \lambda_A H^T i \sigma_2 \Delta^\dagger H S_1^* - \frac{2}{3} \lambda'_{12}S_{1}^* S_{2}^3 +\hc \right]\\
   &\!\!\!\left.\begin{array}{l}
    + \dfrac{\lambda_\Delta}{4}\vev{\bil{\Delta}}^2
    + \dfrac{\lambda'_\Delta}{4}\vev{\bil{\Delta}\bil{\Delta}} +\dfrac{\lambda_1}{4}(\bils{S_1})^2
     + \lambda_{12}(\bils{S_1})(\bils{S_2})\\[0.4cm]
 +\lambda_{H\Delta}(\bil{H})\vev{\bil{\Delta}} + \lambda_{H\Delta}'\vev{H^\dagger \Delta\Delta^\dagger H}
 + \lambda_{1\Delta}\vev{\bil{\Delta}}(\bils{S_1})+ \lambda_{2\Delta}\vev{\bil{\Delta}}(\bils{S_2}),
 \end{array}\qquad\right\} \,\,{\text{ ``incidental''~terms}}\nonumber
\end{align}
\end{widetext}
where the parameters more relevant for the mechanism and the phenomenology are in the first two lines. Although the quartic couplings on 
the third and fourth lines are important for the stability of the potential, they play almost no role otherwise (thus called ``incidental''). The 
stability of the potential is not a primary concern of this manuscript, but it is important to note that the quartic couplings $\lambda_A$ 
and $\lambda_{12}'$ tend to destabilize the potential, and hence are expected to be small. For more considerations regarding stability 
see Appendix~B. We define the neutral components of the fields as 
$H^0=(v+h+i a)/\sqrt{2}$, $\Delta^0=(v_\Delta+\delta+i a_\delta)/\sqrt{2}$ and $S_j=(v_j+s_j+i a_j)/\sqrt{2}$, for $j=1,2$.

The positive mass terms for $\Delta$ and $S_1$ ensure that if $\lambda_A=\lambda_{12}'=0$ then the vevs for these fields are zero. Notice 
that these two quartic couplings are protected from loop corrections by accidental global $U(1)$ symmetries. Moreover, $\lambda_A$ and 
$\lambda_{12}'$ can be made real by rephasing the scalar singlet fields. As long as $v_\Delta$ and $v_1$ are much smaller than $v$ 
and $v_2$, we can obtain the former vevs by treating $H$ and $S_2$ as background fields. First we obtain the approximate vevs of $H$ 
and $S_2$ by setting the other scalar fields to zero, that is,
\begin{equation}
  m_H^2=\frac{1}{2}\lambda_H v^2 + \lambda_{2H}v_2^2,\qquad
  m_2^2=\frac{1}{2}\lambda_2 v_2^2 + \lambda_{2H}v^2.
\end{equation}
Then, by replacing $H$ and $S_2$ by their vevs, we can easily calculate the vevs and the spectrum of the other scalars:
\begin{equation}
  v_1=\frac{\lambda'_{12}v_{2}^3}{3M_1^2},\qquad
  v_\Delta=\frac{\lambda_A v^2 v_1}{2M_\Delta^2},
\end{equation}
and
\begin{subequations}
	\begin{align}  
		M_h^2&=\frac{1}{2}\lambda_Hv^2, \label{eq:mh}\\
  		M_{1}^2&=m_1^2+\frac{1}{2}(\lambda_{1H}v^2+\lambda_{12}v_2^2),\\
 		M_2^2&=\frac{1}{2}\lambda_{2}v_2^2,\\
 		M_\Delta^2&=m_\Delta^2+ \frac{1}{2}\left[\lambda_{2\Delta}v_2^2+(\lambda_{H\Delta}+\lambda_{H\Delta}')v^2\right].\label{eq:mdelta}
	\end{align}
\end{subequations}

The physical masses of the scalars are approximately given by the $M$'s in Eqs.~(\ref{eq:mh}-\ref{eq:mdelta}). Here we see the mechanism 
at work: $\lambda_{12}'$ induces a suppression from $v_2$ to $v_1$, and $\lambda_A$ induces a further suppression from  $v_1$ to 
$v_\Delta$. It is useful to write these quartics in terms of the scalar masses and vevs,
\begin{subequations}
	\begin{align}
  	\lambda_A&=0.008\left(\frac{M_\Delta}{500~\GeV}\right)^2\left(\frac{v_\Delta/\keV}{v_1/\MeV}\right),\\
 	 \lambda_{12}'&=0.03\,\frac{(M_1/100~\GeV)^2(v_1/\MeV)}{(v_2/10~\GeV)^3}.
	\end{align}
\end{subequations}
Note that these relations do not depend on the number of steps, as long as the perturbation theory holds. 

\vskip 0.1cm
\noindent
{\bf  III. Spectrum and mixing phenomenology}\\
The scalar spectrum of this 2-step scenario consists of the 4 aforementioned neutral scalars $(h, \,\delta,\,s_1,\,s_2)$, singly and  doubly 
charged scalars $\delta^+$ and $\delta^{++}$, with masses approximately given by $M_\Delta$, two massive pseudoscalar degrees of 
freedom $(a_\delta,\,a_1)$ with masses approximately given by $M_\Delta$ and $M_1$, and two massless Goldstone bosons. One of the 
Goldstones is the longitudinal polarization of the $Z$ boson while the other one is a massless Majoron, 
$J$~\cite{Chikashige:1980ui, Chikashige:1980qk, Schechter:1981cv}. We will analyze the Majoron phenomenology in the following section.

\begin{table}[t]
  \begin{center}
    \begin{tabular}{|c|c|} \hline
Mixing  & Phenomenology \\ \hline
$h-s_2$   & Higgs observables, direct $s_2$ production \\ \hline
$\delta-s_1$   & New LHC signatures, $s_1$ decay modes \\ \hline
$h-s_1$   & $s_1$ decay modes \\ \hline
$s_1-s_2$   & Irrelevant \\ \hline
    \end{tabular}
  \end{center}	
  \caption{\label{tab:mixings} Sizable scalar mixings  and their phenomenological impact.}
\end{table}

The mixings among the CP even scalars will have important phenomenological impacts (see Table~\ref{tab:mixings} for a summary).
The mixings between $h-s_2$, $\delta-s_1$ and $h-s_1$ are given by
\begin{subequations}
\begin{align}\label{eq:theta_h2}
  \theta_{h2}&\simeq\frac{\lambda_{2H}\,v_2\, v }{M_h^2-M_2^2}
  	\simeq0.16\lambda_{2H}\left(\frac{v_2}{10~\GeV}\right)\beta_{h2},\\
	\label{eq:theta_del1}
  \theta_{\delta1}&\simeq\frac{\lambda_{A}}{2}\frac{v^2}{M_1^2-M_\Delta^2}\simeq10^{-3}\left(\frac{v_\Delta/\keV}{v_1/\MeV}\right)\beta_{1\delta},\\
  \label{eq:theta_h1}
    \theta_{h1}&\simeq \frac{\lambda_{1H} v_1\,v}{M_h^2-M_1^2}
	  	\simeq 1.5 \cdot 10^{-5} \lambda_{1H}\left(\frac{v_1}{\MeV}\right)\beta_{h1},
\end{align}
\end{subequations}
where $\beta_{ab}\equiv (1-M_b^2/M_a^2)^{-1}$.
First, the Higgs mixing with $s_2$ could in principle be sizable. Observations  of Higgs production and decay modes together with 
precision electroweak measurements constrain the mixing angle $\alpha$ with a scalar singlet to be about $\sin\theta_{h2}\lesssim 0.2-0.3$ 
for a 200$-$800~GeV singlet mass~\cite{Robens:2016xkb}. If the scalar is much lighter than the Higgs, for instance in the region 
$1<M_2<10~\GeV$, the constraints on the mixing range from $\sin\theta_{h2}\lesssim 10^{-3}-10^{-1}$~\cite{Clarke:2013aya}.
This Higgs-singlet mixing can lead to very interesting phenomenology, but it is not an exclusive signature of our model.
For small values of $v_2$, the invisible Higgs decay to a pair of Majorons strongly constrains this mixing, as we will see later.

\begin{figure}[t]
\centering
\includegraphics[width=\columnwidth]{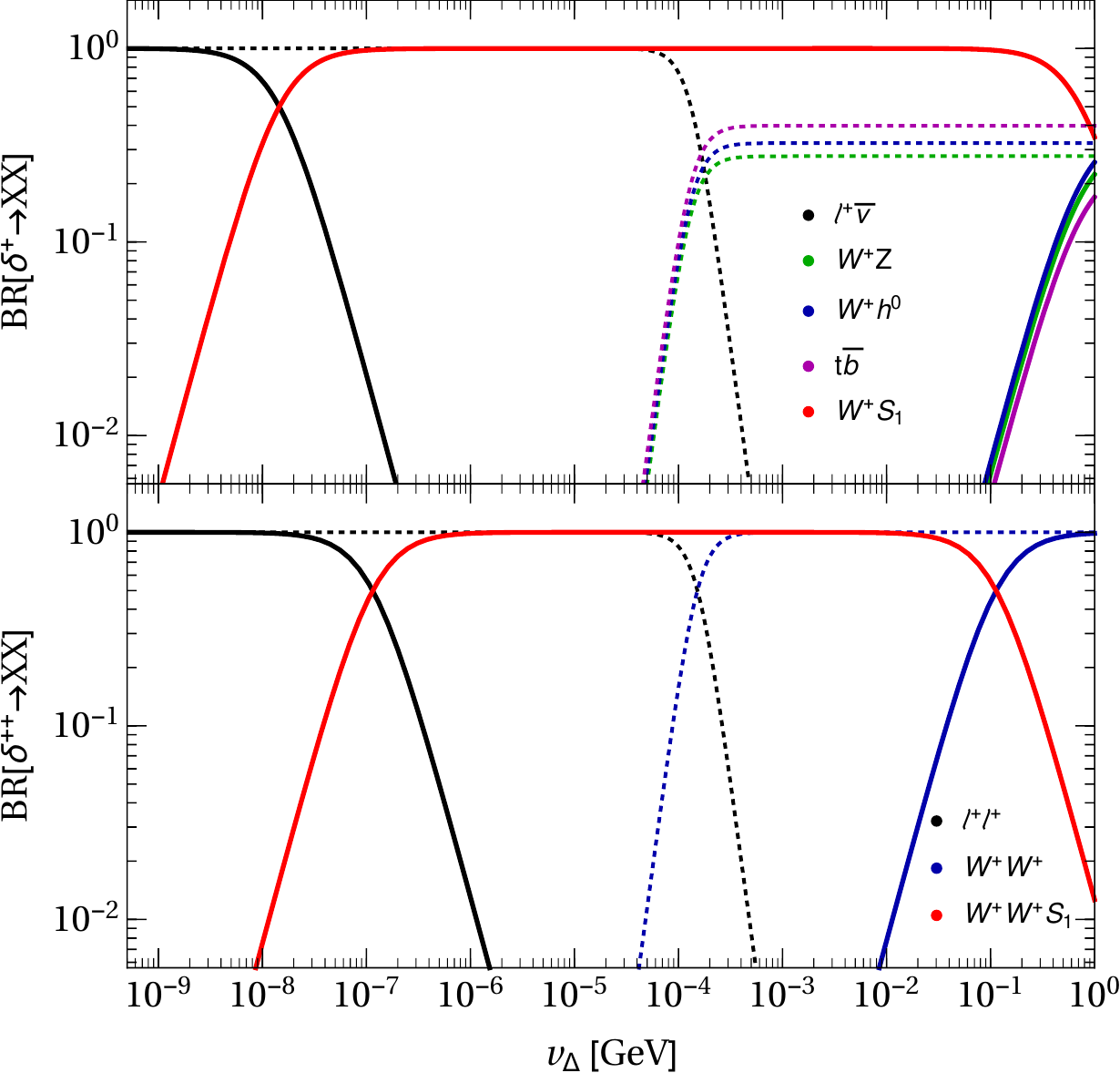}
\caption{Branching ratios of $\delta^+$ (upper panel) and $\delta^{++}$ (lower panel) as a function of the triplet vev $v_\Delta$ for the usual
type II seesaw model (dotted) and our generalized version (solid) . We considered $m_0 = 0.1$ eV, as the lightest neutrino mass, 
$M_{\delta^+} = M_{\delta^{++}} = 500~\GeV$, $M_{1} = 100~\GeV$, and $\theta_{\delta 1} = 0.005$.}
 \label{fig:BRdeltap}
\end{figure}

The mixing between $\delta$ and $s_1$ is quite special, as it leads to drastic deviations from the usual type II seesaw phe\-no\-me\-no\-lo\-gy. 
For $\delta^{++}$, a new decay channel may open up, $\delta^{++}\to W^+ W^+ s_1$ with $s_1$ typically decaying to neutrinos 
(via mi\-xing with $\delta$), quarks or gauge bosons (both via mi\-xing with the Higgs) depending on its mass. Similarly, one can 
have $\delta^+\to W^+ s_1$  and $\delta\to h s_1$. Another distinctive feature is the possibility of having sizable visible decays of 
the pseudoscalar,  $a_\delta\to Z s_1$. Differently from type II seesaw, these decays are controlled uniquely by the gauge coupling 
and the mixing angle $\theta_{\delta1}$. We summarize these features in Table~\ref{tab:decays}. As can be seen in Fig.~\ref{fig:BRdeltap} 
the new decays can dominate a large region of parameter space in the generalized type II seesaw (solid lines) compared to the usual 
case (dotted lines). As we will see later, these new decays provide a smoking gun signature at the LHC, not only opening the possibility 
for discovering the new particles, but also distinguishing the model from type II seesaw.

\begin{table}[t]
  \begin{center}
    \begin{tabular}{|c|c|c|c|} 
\hline
scalar  & Type II & Generalized type II & parameters \\ \hline
$\delta^{++}$ & $\ell^+\ell^+, W^+W^+$ & $W^+W^+ s_1$ & $v_\Delta,\theta_{\delta1}$\\ \hline
$\delta^{+}$ & $\ell^+\nu, W^+\!Z,W^+\!h,\,t\bar b$ & $W^+ s_1$ & $v_\Delta,\theta_{\delta1}$\\ \hline
$\delta$ & $\nu\nu, W^+W^-\!,\,ZZ,hh$ & $h\, s_1$  & $v_\Delta,\theta_{\delta1}$\\ \hline
$a_\delta$ & $\nu\nu,\,t\bar t, Zh$ & $Z\, s_1$ &  $v_\Delta,\theta_{\delta1}$\\ \hline
$s_1$ & not present & $\nu\nu, \,q\bar q, W^+ W^-\!,ZZ$  & $v_\Delta,\theta_{\delta 1},\theta_{h1}$\\ \hline
    \end{tabular}
  \end{center}	
  \caption{\label{tab:decays} Typical decay modes in type II seesaw and new modes in the generalized type II framework. In the last column it is indicated the most relevant parameters governing the partial widths. }
\end{table}

Finally, the mixing between the Higgs and $s_1$ given in Eq.~(\ref{eq:theta_h1}), although small,  plays a significant role in the scalar phenomenology. 
The $s_1$ decay   to charged fermions, driven by $\theta_{h1}$, will compete with the invisible decay to neutrinos, sourced by $\theta_{\delta1}$. 
By analyzing the ratio of these partial widths (see Appendix C for more details),
\begin{equation}
  \frac{\Gamma_{s_1\to\nu\nu}}{\Gamma_{s_1\to ff}}\simeq \frac{3.1}{N_c}\left(\frac{\theta_{\delta 1}/10^{-3}}{\theta_{h1}/10^{-5}}\right)^2 \!\left(\frac{ m_\nu/0.1~\eV}{m_f/\GeV}\right)^2\! \left(\frac{\keV}{v_\Delta}\right)^2\!,\nonumber
\end{equation}
we can see that either visible or invisible $s_1$ decays can dominate in large natural regions of the parameter space. In this manuscript we will focus on the latter. Besides, there is some region of parameter space in which $s_1$ decays to $b$ quarks and  gives rise to displaced vertices at the LHC. We will nevertheless refrain from analyzing that possibility here.

\vskip 0.1cm
\noindent
{\bf IV. Majoron phenomenology}\\
Before dwelling on the LHC signatures, we will first discuss the Majoron phenomenology. Although a massless particle in the spectrum may 
at first seem problematic, its couplings to standard  model fermions are extremely suppressed due to the hierarchy of vevs. The Majoron field 
is the linear combination 
\begin{equation}
  J \simeq \frac{1}{\ell_2 v_2}\left(\ell_1 v_1 a_1 + \ell_2 v_2 a_2 + \frac{1}{2}v_\Delta a_\delta - \frac{v_\Delta^2}{v}a\right),
\end{equation}
where $\ell_{1}=1$ and $\ell_2=1/3$ are the lepton numbers of the corresponding scalars.
It is straightforward to see that the Majoron has very small couplings to charged fermions given by
\begin{align}
  G_{Jff}&= \frac{y_f}{\sqrt{2}} \frac{v_\Delta^2}{\ell_2 v_2 v}=\frac{1.6\cdot 10^{-18}}{\ell_2}\frac{(m_f/\GeV)(v_\Delta/\keV)^2}{(v_2/10~\GeV)},\nonumber\\
  G_{J\nu\nu}&=\sqrt{2}y_\nu \frac{v_\Delta}{\ell_2 v_2}=  \frac{5\cdot 10^{-12}}{\ell_2}\frac{(m_\nu/0.1~\eV)}{(v_2/10~\GeV)},
\end{align}
easily avoiding constraints from neutrinoless double beta decay with Majoron emission $G_{J\nu\nu}<(0.8-1.6)\times 10^{-5}$~\cite{Gando:2012pj}, 
as well as astrophysical bounds $G_{Jee}<4.3\times 10^{-13}$~\cite{Patrignani:2016xqp}. Although a thermalized Majoron would contribute to 
increase the effective number of relativistic degrees of freedom by $4/7$, the tiny coupling in this scenario leads to very little Majoron production 
in the early universe. 

A stringent bound on the Higgs-$s_2$ mixing comes from Higgs decaying invisibly to a pair of Majorons~\cite{Shrock:1982kd}. It is straightforward to obtain the approximate constraint (see e.g. Ref.~\cite{Dobrescu:2000yn}),
\begin{equation}
  \theta_{h2}<1.5\cdot 10^{-3}\bigg[\frac{v_2}{10~\GeV}\bigg]\left[\frac{\Gamma_h}{4.2~\MeV}\frac{{\rm BR}_{h\to inv}}{0.22}\right]^{1/2}
\end{equation}
where $\Gamma_h$ is the Higgs total width and ${\rm BR}_{h\to{\rm inv}}$ is its in\-vi\-si\-ble branching ratio. The Higgs total width 
has only been measured indirectly, via comparison between on-shell and off-shell Higgs production, yielding the model-dependent bound 
$\Gamma_h^{\rm exp}<13~\MeV$ at 95\% C.L.~\cite{Khachatryan:2016ctc}. The Higgs invisible branching ratio has been bounded to 
be below 0.22~\cite{CMS-PAS-HIG-17-031, Aad:2015pla}. This strong bound on $\theta_{h2}$ could be alleviated by raising $v_2$ to the TeV.

\vskip 0.1cm
\noindent
{\bf V. Collider phenomenology}\\
In this section, we study the collider phenomenology for the generalised type II seesaw model. The leading production channels for this 
framework remain the same as in the usual type II, \ie, the charged Higgs states will be dominantly produced in pairs  via $s$-channel 
electroweak boson exchange, leading primarily to associated production of double and single charged Higgs bosons 
$\delta^{\pm\pm}\delta^{\mp}$, followed by double charged Higgs pair production 
$\delta^{++}\delta^{--}$\footnote{We have checked that producing one triplet scalar in association with $s_1$ is typically sub-leading, 
as it is suppressed by the small mixing $\theta_{\delta1}$. Thus, these production modes will be disregarded here.}. Although these two 
production channels do not present differences in rate between the standard type II seesaw and our new model construction, their 
corresponding decays display new relevant phenomenological signatures. The $\delta$ -- $s_1 $ mixing engenders new interaction 
terms from the triplet kinetic term
\begin{align}
 \mathcal{L}\supset {\rm Tr}[(D_\mu\Delta)^{\dagger}D_\mu\Delta]\,,
 \label{eq:delta}
 \end{align}
making the decays ${\delta^{\pm\pm}\rightarrow W^\pm W^\pm s_1}$  and ${\delta^\pm\rightarrow W^\pm s_1}$ available. 
Note that these  partial widths do not present any $v_\Delta$  suppression, instead  it depends only on gauge couplings, being equally 
large in a wide range of parameter space  $v_\Delta\sim 10^{-7}-10^{-1}~\GeV$, distinctly from the usual type II,  see Fig.~\ref{fig:BRdeltap}.

Therefore, the $pp\rightarrow \delta^{\pm\pm} \delta^{\mp}$ production channel not only reveals the triplet structure nature of 
$\delta^{\pm\pm}$ and $\delta^{\pm}$~\cite{Perez:2008ha, Perez:2008zc}, but can also differentiate our construction from the usual 
type II model. To explore this phenomenology, we analyse the ${pp\rightarrow \delta^{\pm\pm} \delta^{\mp}}$ production at 
the $\sqrt{s}=13$~TeV LHC, focusing on the trilepton plus missing energy signature, with two same flavor and same sign leptons, 
$e^\pm e^\pm\mu^\mp+\slashed{E}_T$ and $\mu^\pm \mu^\pm e^\mp + \slashed{E}_T$. The leptons arise from the $W$-boson
 decays and relevant extra sources of missing energy follow from the dominant $s_1$ decay, $s_1\rightarrow \nu\bar{\nu}$.

\begin{figure}[t]
\centering
\includegraphics[width=0.85\columnwidth]{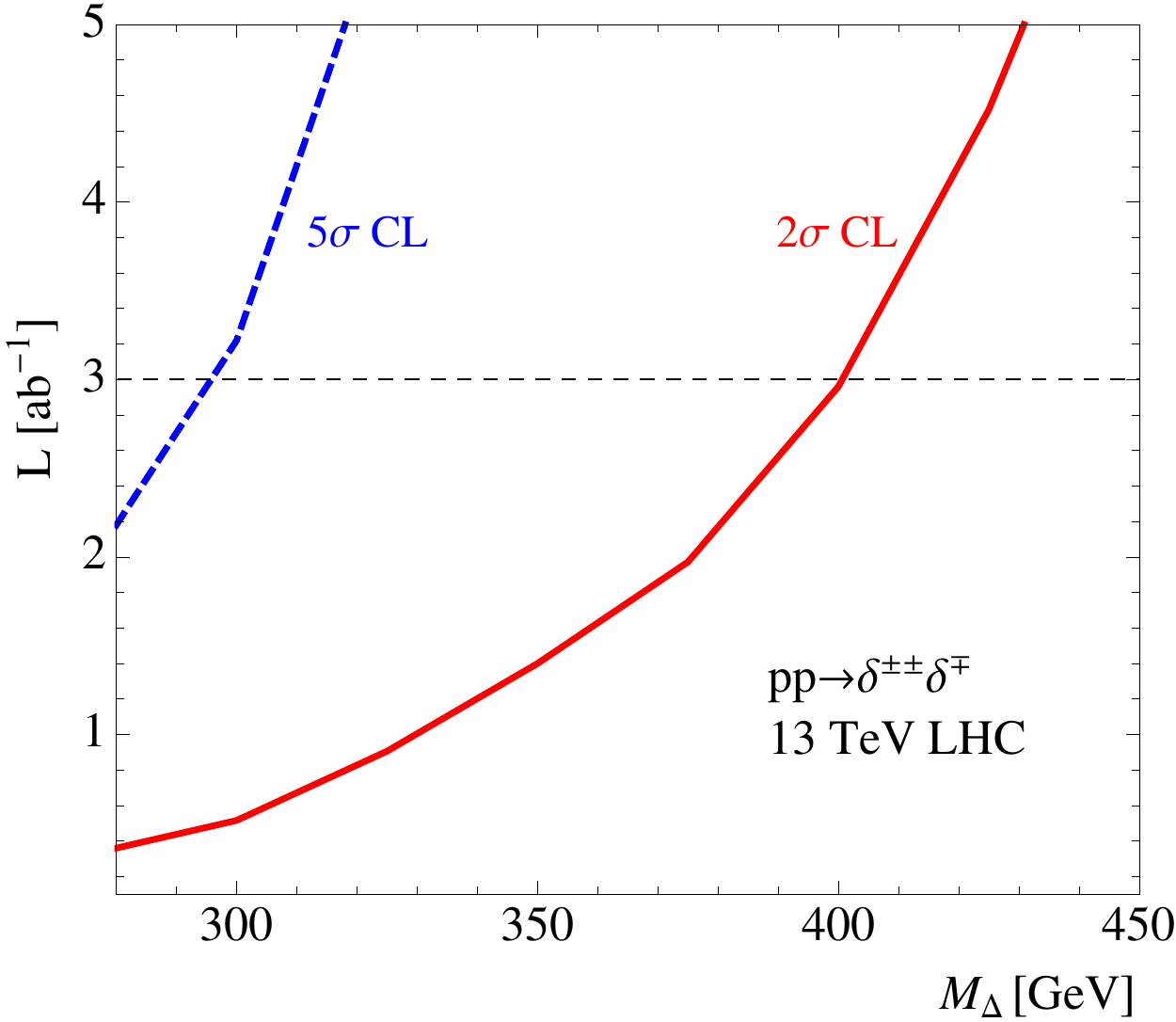}
\caption{Luminosity required to observe ${pp\rightarrow \delta^{\pm\pm}{\delta^{\mp}}}$ as a function of $M_{\Delta}$ at 
$2\sigma$ (red full) and $5\sigma$ (blue dashed) confidence level.
 We assume $M_{1}=100$~GeV and $v_{\Delta}=10^{-6}$~GeV.}
 \label{fig:bound}
\end{figure}

Our model is implemented in FeynRules~\cite{Alloul:2013bka} and the signal sample is generated with MadGraph5~\cite{Alwall:2014hca}. 
A Next-to-leading order QCD K-factor of  1.25 has been applied~\cite{Akeroyd:2005gt}. To obtain a robust simulation of the background 
components, that display large fake rates, our simulation follows the recent 13~TeV CMS study~\cite{Sirunyan:2018mtv}. Although CMS 
targets a heavy neutral Majorana lepton $N$, it presents a set of search regions for the high mass regime $m_N>m_W$, leading to a more 
sizable $\slashed{E}_T$, that also applies to our model.

In this analysis, jets are defined with the anti-$k_T$ clus\-te\-ring algorithm with $R=0.4$, $p_{Tj}>25$~GeV and $|\eta_j|<2.4$ via 
Fastjet~\cite{Cacciari:2011ma}. Events with one or more $b$-jets are vetoed with 70\% b-tagging efficiency and 1\% mistag rate. Electrons 
and muons are defined with $|\eta_\ell|<2.4$ and the three leptons must satisfy $p_{T\ell}>55,15,10$~GeV. Finally, the events are 
divided in bins associated to three observables: $i)$ the trilepton mass system $m_{3\ell}$; $ii)$ minimum invariant mass of all opposite 
sign leptons $m_{2\ell OS'}^{min}$; and $iii)$ transverse mass $m_{T}=\sqrt{2p_{T\ell}\slashed{E}_T(1-\cos\phi)}$, where $p_{T\ell}$ 
corresponds to the lepton which is not used in the $m_{2\ell OS'}^{min}$ calculation and  $\phi$ is the azimuthal angle between 
$\vec{p}_{T\ell}$ and $\vec{p}_{T}^{miss}$.

Using the CMS background estimate, we perfom a binned log-likelihood analysis based on the CL$_s$ method~\cite{Read:2002hq}, 
exploring all search regions with $e^\pm e^\pm\mu^\mp+\slashed{E}_T$ and $\mu^\pm \mu^\pm e^\mp + \slashed{E}_T$ 
displayed by Ref.~\cite{Sirunyan:2018mtv}.  In Fig.~\ref{fig:bound}, we present the luminosity required to observe 
${pp\rightarrow \delta^{\pm\pm}{\delta^{\mp}}}$ as a function of $M_{\Delta}$ at $2\sigma$ (red full) and $5\sigma$ (blue dashed) 
confidence level. At the high-luminositiy LHC, $\mathcal{L}=3$~ab$^{-1}$, we can discover charged Higgses at $5\sigma$ level  up
 to $M_{\Delta}=300$~GeV  and exclude it at $2\sigma$ level  up to $M_{\Delta}=400$~GeV.

\vskip 0.1cm

A final comment is in order regarding two phenomenological aspects beyond the ones discussed so far. First, our model may also induce 
lepton flavor violation processes,  very similar to the usual type II seesaw scenario~\cite{Dinh:2012bp}. Second, although the model does 
not have enough CP violation, adding a se\-cond $SU(2)$ triplet scalar~\cite{Ma:1998dx} may lead to successful leptogenesis. The study of 
such possibilities is beyond the scope of this manuscript.

\vskip 0.1cm
\noindent
{\bf VI. Conclusions}\\
In this Letter we have proposed a generalization of type II seesaw in which  lepton number is broken dynamically and no hierarchy 
problem neither arbitrarily small parameters are present. The rich phenomenology of the model includes de\-via\-tions of standard Higgs 
couplings, the presence of a massless neutral pseudoscalar and  more importantly a novel smoking gun signature at the LHC. This 
distinctive new signature may reveal the triplet nature of the charged scalars and at the same time  disentangle the framework from 
the usual type II seesaw model.

\vspace{0.1cm}

\begin{acknowledgments}{\it Acknowledgments}  We thank K. Agashe, E. Bertuzzo and R. Zukanovich Funchal for useful discussions. 
We are grateful to K. Babu for careful reading of the manuscript. Fermilab is operated by Fermi Research Alliance, LLC, under Contract 
No. DE-AC02-07CH11359 with the US Department of Energy.  JG and PM acknowledge support from the EU 
grants H2020-MSCA-ITN-2015/674896-Elusives and H2020-MSCA-2015-690575-InvisiblesPlus. DG was funded by U.S. National Science 
Foundation under the grant PHY-1519175. YP ack\-now\-led\-ges support from Funda\c{c}\~{a}o de Amparo \`a Pesquisa do Estado 
de S\~{a}o Paulo (FAPESP), under processes 2013/03132-5 and 2017/19765-8, and from Conselho Nacional de Desenvolvimento 
Cient\'{i}fico e Tecnol\'{o}gico (CNPq).

\end{acknowledgments} \medskip

\appendix
\section{I. Supplemental Material}
\subsection{A. \emph{n}-step generalized type II seesaw}
\label{app:n-step}

Here we present the generalization of our framework for an arbitrary number of scalar singlets $n$. We define the following scalar bilinears,
\begin{equation}
  B_i\equiv S_i^* S_i,\quad B_\Delta \equiv \tr(\Delta^\dagger\Delta),\quad B_H\equiv H^\dagger H,
\end{equation}
which allow to write the  scalar potential in a compact form
\begin{align}
  &V= -\frac{m_H^2}{2}B_H  +\sum_\varphi^{\Delta,1..n-1}m_\varphi^2B_\varphi - \frac{m_n^2}{2}B_n+ \sum_\varphi^{\rm all}\frac{\lambda_\varphi}{4}B_\varphi^2   \nonumber\\
  & + \!\!\!\sum_{\varphi,\varphi'>\varphi}^{\rm all}\!\! \lambda_{\varphi\varphi'}B_\varphi B_{\varphi'} +\frac{\lambda'_\Delta}{4}\tr(\Delta^\dagger\Delta\Delta^\dagger\Delta) + \lambda'_{H\Delta}H^\dagger\Delta\Delta^\dagger H\nonumber\\
  & + \left[\lambda_A H^T i \sigma_2 \Delta^\dagger H S_1^*  - \frac{2}{3}\sum_{i=1}^{n-1}\lambda'_{i,i+1}S_i^* S_{i+1}^3
  	 + \hc\right].
\end{align}
The notation in the sum  of the first term of the second line indicates that permutations of $\lambda_{\varphi\varphi'}$ should not be taken 
(to avoid double counting). Without loss of generality, all $\lambda_{i,i+1}'$ and $\lambda_A$ can be made real by rephasing the scalar 
singlet fields. The masses and vevs in the $n$-step realization are approximately given by
\begin{subequations}
\begin{align}
  m_H^2&=\frac{1}{2}\lambda_H v^2 + \lambda_{nH}v_n^2,\\
  m_n^2&=\frac{1}{2}\lambda_n v_n^2 + \lambda_{nH}v^2,\\
  v_i&=\frac{\lambda'_{i,i+1}v_{i+1}^3}{3M_i^2},\quad{\rm for}~i=1,\dots,n-1,\\
  v_\Delta&=\frac{\lambda_A v^2 v_1}{2M_\Delta^2},\\
  M_h^2&=\frac{1}{2}\lambda_Hv^2,\\
  M_{i}^2&=m_i^2+\frac{1}{2}(\lambda_{iH}v^2+\lambda_{in}v_n^2), \quad i=1,\dots,n-1,\\
  M_n^2&=\frac{1}{2}\lambda_{n}v_n^2,\\
  M_\Delta^2&=m_\Delta^2+ \frac{1}{2}\left[\lambda_{n\Delta}v_n^2+(\lambda_{H\Delta}+\lambda_{H\Delta}')v^2\right].
\end{align}
\end{subequations}
These expressions should hold in the regime, $v_i\ll v_{i+1}$, that is, \begin{equation}\label{eq:perturbation}
\varepsilon\equiv\lambda_{i,i+1}' \frac{v_{i+1}^2}{3M_i^2}\ll 1.
\end{equation}

In fact, it is straightforward to show that as long as Eq.~(\ref{eq:perturbation}) is satisfied, for any number $n$ of scalar singlet fields, 
the vev of $s_j$, $j<n$, is simply given by
\begin{equation}
  v_j=\prod_{k=0}^{n-j-1}\left(\frac{\lambda'_{j+k,j+k+1}}{3}\frac{v_n^2}{M_{j+k}^2}\right)^{3^{k}}v_n.
\end{equation}
If, for simplicity, one takes all $\lambda'_{ij}=\lambda'$ and $M_i=M$, then we obtain a simplified expression,
\begin{equation}
  v_j = \left(\frac{\lambda'}{3}\frac{v_n^2}{M^2}\right)^K v_n,\quad K=(3^{n-j}-1)/2.
\end{equation}
We can clearly identify the parametric suppression $\varepsilon^K$ res\-pon\-si\-ble for  making $v_1\ll v_n$. For instance, 
if $\varepsilon=0.01$ and $n=3$ we obtain $v_1 \sim 10^{-8}v_n$. Note that the expressions for the mixing angles defined in Eqs.~(\ref{eq:theta_h2}-\ref{eq:theta_h1}) are valid for any $n$, and thus the phenomenological considerations re\-gar\-ding 
Higgs couplings, Majoron physics and LHC signatures will still apply.

\subsection{B. Stability of the scalar potential}
Although a complete study on the stability of the scalar potential are not the main focus of this Letter, we provide here sufficient 
conditions for the stability. The key point is that the quartic couplings $\lambda_A$ and $\lambda'_{12}$ (or any $\lambda'_{i,i+1}$ in 
the $n$-step scenario) can always yield negative contributions to the potential when the values of the fields go to infinity, independently 
of their sign. As these couplings are the core of the generalized type II seesaw mechanism, it is  important to understand how to control 
these contribution so that the potential is bounded from below. Although a full analysis of the stability would be very complicated, specially 
in the $n$-step scenario, we can still derive useful sufficient conditions to have stability. The idea is to split the scalar potential into pieces 
that will isolate each $\lambda'_{12}$ or $\lambda_A$,
\begin{equation}
  V =  V_A + V_{12} + \ldots + V_0
\end{equation}
and require each piece to be independently positive. For now we will focus on $n\!=\!2$-steps and generalize the method in the end.

The first piece deals with $\lambda_A$. We define
\begin{align}
  V_A& \equiv \lambda_{1H}(\bils{S_1})(\bil{H}) +\lambda_{1\Delta}\vev{\bil{\Delta}}(\bils{S_1})\\
  &+ \lambda_{H\Delta}(\bil{H})\vev{\bil{\Delta}}+ \left( \lambda_A H^T i \sigma_2 \Delta^\dagger H S_1^* +\hc \right)\nonumber
\end{align}
and require it to be positive. By performing an $SU(2)$ rotation on the field one can always write~\cite{Babu:2016gpg}
\begin{equation}
  i\sigma_2\Delta=\left(
  \begin{array}{c c}
    a & 0 \\
    0 & b \,e^{i\alpha} \\
  \end{array}\right),\quad 
  H = \left(
  \begin{array}{c c}
    c\,e^{i\beta} \\
    d \,e^{i\gamma} \\
  \end{array}\right),
\end{equation}
and $S_i=R_i e^{i\phi_i}$.
Then, it is straightforward to obtain
\begin{subequations}\label{eq:lambda-A}
\begin{align}
  &\lambda_{H\Delta}>0,& &\lambda_{1\Delta}>0,\\
  &\lambda_{1H}>0,&  &|\lambda_A|^2<\lambda_{1H}\lambda_{H\Delta}.
\end{align}
\end{subequations}
Now, we handle $\lambda'_{12}$ by defining
\begin{align}
  V_{12}& \equiv \frac{\lambda_2}{4}(\bils{S_2})^2 + \lambda_{12}(\bils{S_1})(\bils{S_2})\notag\\
  &\qquad\qquad\qquad-\left(\frac{2}{3} \lambda'_{12}S_{1}^* S_{2}^3 +\hc \right),
\end{align}
and requiring $V_{12}>0$. This yields
\begin{equation} \label{eq:lambda-12}
  \lambda_{12}>0,\quad\lambda_2>0,\quad|\lambda'_{12}|^2<\frac{9}{16}\lambda_{12}\lambda_2.
\end{equation}
We still have to deal with seven quartic couplings. First note that $\lambda_{1}$, $\lambda_{2H}$, and $\lambda_{2\Delta}$ need 
to be positive, as there is no other quartic left that can compensate for a negative contribution to the potential sourced by these couplings.
 The remaining parameters, $\lambda_\Delta$, $\lambda'_\Delta$, $\lambda_{H\Delta}$ and $\lambda'_{H\Delta}$, essentially 
 define a usual type II seesaw potential and the stability conditions for that case are known~\cite{Babu:2016gpg}. The requirements 
 for these seven quartics can be summarized as
\begin{subequations}\label{eq:stability-7}
\begin{align}
  &{\rm (i)}\quad\lambda_H>0,\,\,\lambda_{1}>0,\,\,\lambda_{2H}>0,\,\, \lambda_{2\Delta}>0,\\
  &{\rm (ii)} \quad\lambda_\Delta+\lambda'_\Delta>0,\,\,2\lambda_\Delta+\lambda'_\Delta>0,\\
  &{\rm (iii)}\quad2 \lambda'_{H\Delta}+\sqrt{\lambda_H(\lambda_\Delta+\lambda'_\Delta)} >0,\\
  & {\rm (iv)}\quad 2\lambda'_{H\Delta}\sqrt{\lambda_\Delta+\lambda'_\Delta}+(2\lambda_\Delta+\lambda'_\Delta)\sqrt{\lambda_H}>0.
\end{align}
\end{subequations}
We emphasize that if inequalities (\ref{eq:lambda-A}), (\ref{eq:lambda-12}), and (\ref{eq:stability-7}) are all satisfied, then the potential is stable.

The generalization to more $n$-steps is now straightforward. By defining 
\begin{align}
  V_{i,i+1}&\equiv \frac{\lambda_{i+1}}{4}(S_{i+1}^*S_{i+1})^2 + \lambda_{i,i+1}(S_{i}^*S_{i})(S_{i+1}^*S_{i+1})\nonumber\\
    &\qquad\qquad\qquad- \left(\frac{2}{3} \lambda'_{i,i+1}S_{i}^* S_{i+1}^3 +\hc \right),
\end{align}
and requiring $V_{i,i+1}>0$ we obtain 
\begin{equation}
  \lambda_{i,i+1}>0,\quad\lambda_{i+1}>0,\quad|\lambda'_{i,i+1}|^2<\frac{9}{16}\lambda_{i,i+1}\lambda_{i+1}
\end{equation}
for $i=1..n-1$. Again, there are no quartic couplings left to compensate for $\lambda_{iH}$ or $\lambda_{i\Delta}$, which demands
\begin{equation}
    \lambda_{i}>0,\,\,\lambda_{iH}>0,\,\, \lambda_{i\Delta}>0, \quad i=1..n.
\end{equation}
These conditions are by no means necessary, but only sufficient for having stability in the $n$-step realization.

\subsection{C. Partial widths}
We present in this Appendix the partial widths for the novel decay channels of some of the extra scalars in the generalized type II 
seesaw framework. In the case of $\delta$, we will have three new channels: $\delta\to h s_1$, $\delta\to h h s_1$, and 
$\delta\to h s_1 s_1$. As the latter is suppressed by $v_1^2$, we will safely neglect it in the remainder. 
The partial widths for the first two channels are
\begin{align*}
	&\Gamma(\delta\to h s_1) \simeq \frac{v^2}{1024\pi M_{\Delta}}(8\lambda_A\cos(2\theta_{\delta 1})-\lambda_{1H}\sin(2\theta_{\delta 1}))^2,\\
	&\Gamma(\delta\to h h s_1) \simeq \frac{M_{\Delta}}{8192\pi^3}(2\lambda_A\cos(2\theta_{\delta 1})-\lambda_{1H}\sin(2\theta_{\delta 1}))^2 ,
\end{align*}
where we have neglected the phase space factor by assuming $M_1+2M_h\ll M_\Delta$.
The phase space for 2-body decay can easily be incorporated by multiplying the partial width by 
\begin{equation}
	\bar{\beta}_{\delta\to h s_1}\equiv\sqrt{1-\frac{2(M_1^2+M_h^2)}{M_{\Delta}^2}+\frac{(M_1^2-M_h^2)^2}{M_{\Delta}^4}}	.
\end{equation}
The decay width ratios with respect to the leptonic channel, $\delta\to \nu \nu + \bar{\nu} \bar{\nu}$, are approximately given by
\begin{align*}
	\frac{\Gamma[\delta\to h s_1]}{\Gamma[\delta\to \nu \nu + \bar{\nu} \bar{\nu}]}&\simeq  \lambda_A^2\frac{v_\Delta^2}{\sum_i m_{\nu_i}^2}\frac{v^2}{M_{\delta}^2},\\
	\frac{\Gamma[\delta\to h h s_1]}{\Gamma[\delta\to \nu \nu + \bar{\nu} \bar{\nu}]}&\simeq  \frac{\lambda_A^2}{512\pi^2}\frac{v_\Delta^2}{\sum_i m_{\nu_i}^2}.
\end{align*}

In the case of the single-charged scalar $\delta^+$, the additional channel $\delta^+\to W^+ s_1$ is the most relevant.
Its decay width is given by
\begin{align*}
	\Gamma[\delta^+\to W^+ s_1] &= \cos^2\eta\,\frac{\sin^2(\theta_{\delta 1})}{8\pi}\frac{M_{\delta^+}^3}{v^2}\bar{\beta}_{\delta^+}^3,
\end{align*}
with
\begin{align*}
\cos^2\eta &\equiv 1-\frac{2 v_\Delta^2}{v^2},\\
\bar{\beta}_{\delta^+}&\equiv\sqrt{1-\frac{2(M_1^2+M_W^2)}{M_{\delta^+}^2}+\frac{(M_1^2-M_W^2)^2}{M_{\delta^+}^4}}	.
\end{align*}
The ratio with the leptonic channel is approximately
\begin{align*}
	\frac{\Gamma[\delta^+\to W^+ s_1]}{\Gamma[\delta^+\to \ell^+ \nu]}&\simeq  2\sin^2(\theta_{\delta 1})\frac{v_\Delta^2}{m_{\nu_i}^2}\frac{M_{\delta^+}^2}{v^2}.
\end{align*}

For $s_1$, we have the decay into charged fermions and neutrinos
\begin{subequations}
\begin{align}
  \Gamma[s_1\to f\bar f]&=N_c\frac{M_1}{8\pi}\frac{m_f^2}{v^2}\sin^2\theta_{h1}\bar\beta_{1},\\
  \Gamma[s_1\to \nu\nu]&=\frac{M_1}{16\pi}\frac{m_\nu^2}{v_\Delta^2}\sin^2\theta_{\delta1},
\end{align}
\end{subequations}
where $N_c$ is the number of colors and $\bar\beta_{1}\equiv(1-4m_f^2/M_1^2)^{3/2}$.
We do not present the analytic expressions for the new 3-body decay channel $\delta^{++}\to W^+ W^+ s_1$, as it is not particularly illuminating. 

\bibliography{Bib}

\appendix
\setcounter{equation}{0}
\setcounter{figure}{0}
\setcounter{table}{0}
\setcounter{section}{0}
\makeatletter
\renewcommand{\theequation}{A\arabic{equation}}
\renewcommand{\thefigure}{A\arabic{figure}}
\renewcommand{\thetable}{A\arabic{table}}

\end{document}